\documentclass[11pt]{article}
\usepackage{array}
\usepackage{graphicx}
\usepackage{textpos}

\title{Superheavy Light Quarks and the Strong P, T Problem}

\author{Frank Wilczek\\
\small\it Center for Theoretical Physics, MIT, Cambridge MA 02139 USA \\
\small\it Origins Project, Arizona State University, Tempe AZ 25287
USA \\
Guy D.\ Moore \\
  \small\it Technische Universit\"at Darmstadt,
  64289 Darmstadt Germany}

\begin{document}

\maketitle

\begin{textblock*}{5cm}(11cm,-8.2cm)
\fbox{\footnotesize MIT-CTP-4759}
\end{textblock*}

\begin{abstract}
New superstrong forces, analogous to QCD but featuring a larger mass
scale, should they exist, offer new perspectives on the strong P, T
problem.   If the superstrong dynamics supports confinement without
chiral symmetry breaking, then one might implement the ``massless
quark'' solution in a phenomenologically acceptable way, by using a
massless quark that is always confined within superheavy particles,
and is therefore effectively superheavy: a cryptoquark.  Assuming
confinement and chiral symmetry breaking from the superstrong
dynamics, we find a new mechanism to generate an axion field without
introducing new fundamental scalar fields.
%
\end{abstract}

\medskip

\bigskip

Our present-day core theories -- general relativity, and the
$SU(3)\times SU(2) \times U(1)$ gauge theories of strong, weak, and
electromagnetic interactions -- offer a profound explanation of the
origin of (approximate) invariance of fundamental physics under the
time reversal operation $T$.   As emphasized by Kobayashi and Maskawa
\cite{kobayashi}, (approximate) $T$ symmetry arises as an accidental
consequence of more basic principles, i.e. the principles of
relativistic quantum field theory, together with the observed
multiplet structure of matter in $SU(3) \times SU(2) \times U(1)$.
Since $T$ symmetry is emphatically {\it not\/} manifested in
macroscopic observables, it might otherwise appear as a gratuitous and
even problematic feature of our world-description.

At first sight the three-family version of $SU(3)\times SU(2) \times
U(1)$ appears to allow only a single $T$-violating parameter.  That
parameter is a weak mixing angle involving transitions to the third
generation.  The effects it induces are predicted to be visible only
in $K - \bar K$ mixing and in processes involving the third
generation, and in fact they have been observed.

There is, however, a feature of that explanation which remains
unsatisfactory.   Deeper investigation reveals an additional
possibility for $T$ violation, bringing in an independent parameter
\cite{thooft}.   The additional parameter is the coefficient of a
color gauge field version of $\Delta {\cal L} \propto \vec E \cdot
\vec B$, which can appear as an interaction term in the theory's
Lagrangian density.   Since that interaction is a total divergence, it
does not appear in the classical equations of motion.  But the
quantity of which it is a divergence is not gauge invariant, so in the
quantum theory it can (and does) exhibit singular fluctuations,
leading to surface terms.  Thus the $\vec E \cdot \vec B$ interaction
term can (and does) influence the quantum theory.

One can parameterize the coefficient of the new interaction in terms
of a variable $\theta$, which is periodic modulo 2$\pi$.  $\theta$ is
odd under $T$ (and $P$), so those symmetries are broken unless
$$
\theta ~\equiv~ 0, \pi  \ \  ({\rm mod} \  2 \pi)
$$ 
The new interaction can be calculated to induce an electric dipole
moment for the neutron, which violates both $P$ and $T$ invariance.
No such moment has been observed, despite very sensitive searches.
That result, together with some special considerations that exclude
$\theta \approx \pi$, lead to the bound \cite{edmBound}
$$
| \theta | ~<~ 10^{-10} \ \ ({\rm mod} \  2 \pi)
$$
The smallness of this number does not follow from general principles.
It constitutes a remaining gap in our understanding, often called the
strong $P, T$ problem.

\begin{figure}[h]
\centering
\includegraphics[scale=0.52]{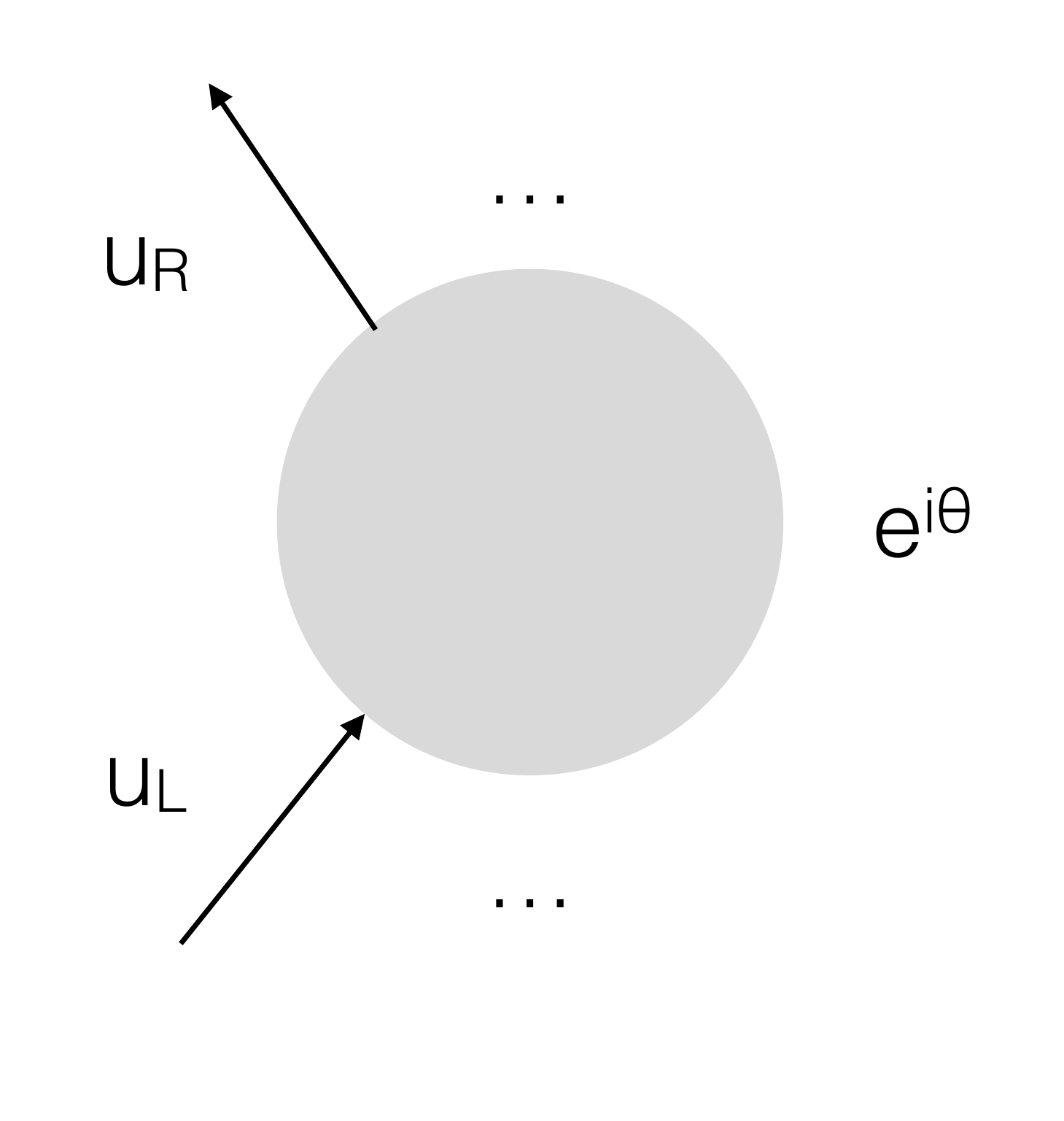}\\
\caption{Gauge configurations with unit topological charge are
  weighted with the phase parameter $e^{i\theta}$.  In the presence of
  such gauge fields charged fermion fields have zero modes whose
  quantum numbers, including their chiral structure, implement
  appropriate anomaly equations (Atiyah-Singer index theorem).  The
  ellipsis indicates additional quark species, beyond
  $u$.}\label{thooftVerticesOne}
\end{figure}

One can visualize the role of $\theta$ by introducing an effective
interaction, the 't Hooft vertex, depicted in Figure
\ref{thooftVerticesOne}.    The blob in the 't Hooft vertex indicates
a gauge field configuration with unit topological charge, and the
fermion lines emanating from it indicate fermion emission and
absorption.   One left-handed quark of each flavor is absorbed, and
one right-handed quark of each flavor is emitted.   (This
representation is schematic, but adequate to our purposes.)   The
interaction contains an overall factor $e^{i \theta}$.    Now if the
$u$ quark were massless, then it would be possible to redefine $u_L$
by a phase factor without changing the form of any term in the
Lagrangian other than the 't Hooft vertex.  In particular, we could
use that freedom to absorb the factor $e^{ i \theta}$ into a
re-definition of $u_L$.  In the new variables we would have $\theta =
0$.  Thus the standard $P$ and $T$ operations would become manifest
symmetries of the effective interaction, and the strong $P, T$ problem
would be solved.

This proposed ``zero mass'' solution is almost as old as the strong
$P, T$ problem itself.  (See, for example, \cite{fw}.)  But it appears
that in reality $m_q \neq 0$ for every one of the known quarks,
implying that Nature has not chosen to use that solution
\cite{quarkMasses}.


It is logically possible, however, that there are massless (either
literally, or in a sense to be defined below) ``quarks'' $Q$ -- that
is, spin $\frac{1}{2}$ fermions carrying non-trivial color quantum
numbers -- yet to be discovered.   It will be convenient to have a
name for such particles, and we shall call them cryptoquarks.  Indeed,
there is no necessary relation between the current mass of confined quarks and
the mass of the lightest particles which contain them.   If there is a
new strong interaction $G$, with dynamics resembling that of QCD's
$SU(3)$  but on a much larger mass scale, and massless spin
$\frac{1}{2}$ fermions carrying both $G$ and $SU(3)$ quantum numbers,
then direct evidence for the existence of those fermions (or, for that
matter, of $G$) could be difficult to obtain.   One must exceed the
threshold for production of the (encrypting) particles containing the
cryptoquarks, or the crossover, at higher energy, to liberation of $G$
jets.  (Related ideas are expressed in \cite{appelquist}
\cite{hsuSannino}.)

\begin{figure}[t]
\includegraphics[scale=0.42]{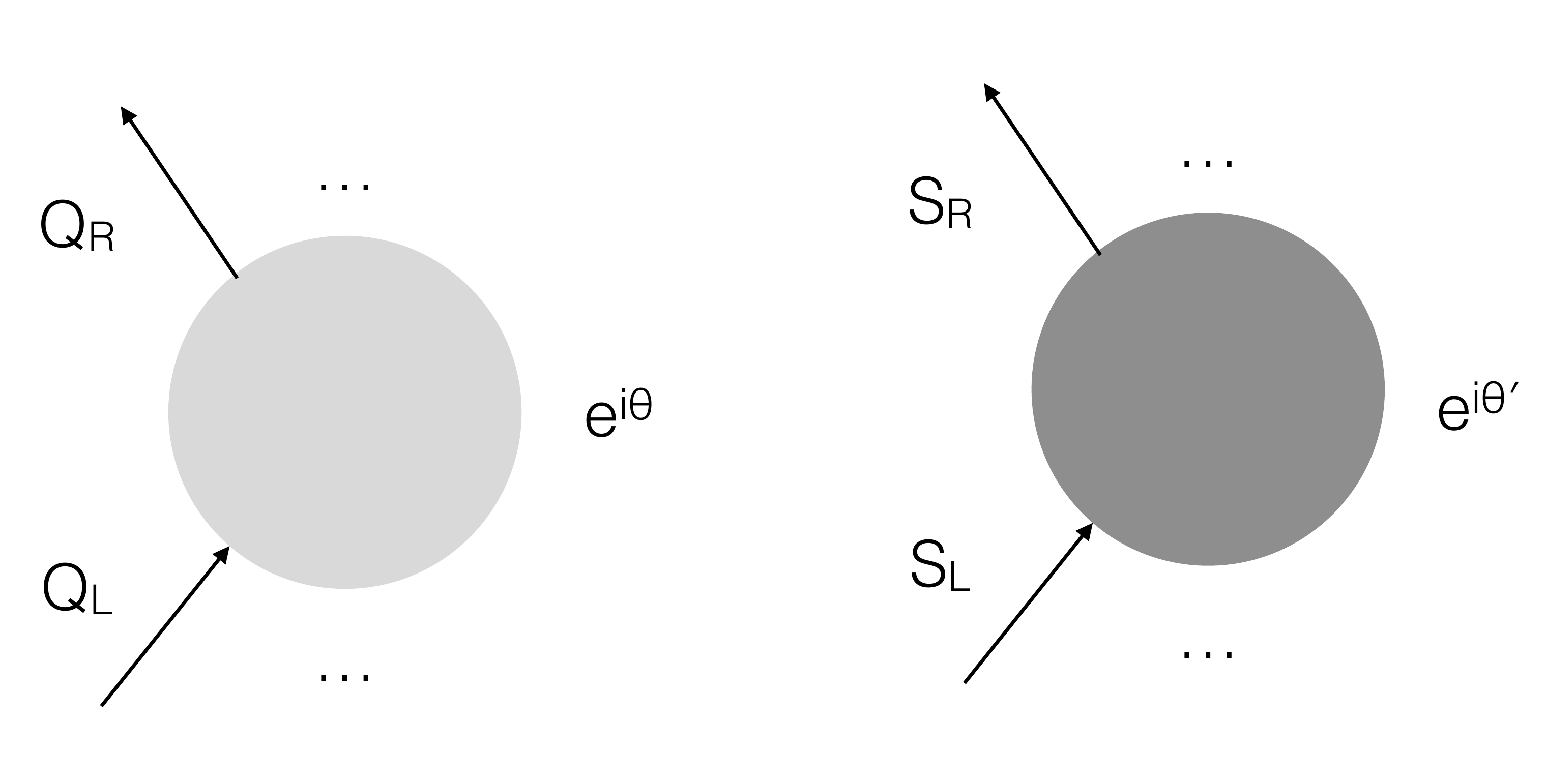}\\
\caption{On the left, the 't Hooft vertex for color $SU(3)$, with the
  presence of $Q$ highlighted.  The ellipsis indicates other quark
  species, possibly including additional samples of $Q$, and of $S$.
  On the right, the 't Hooft vertex for a hypothetical new superstrong
  interaction with gauge group $G$.  Gauge configurations with unit
  topological charge are weighted with the phase parameter
  $e^{i\theta^\prime}$.  The ellipsis indicates additional quark
  species, possibly containing additional samples of $S$ and of
  $Q$.}\label{thooftVerticesTwo}
\end{figure}

The mechanism whereby cryptoquarks solve the strong $P, T$ problem is
essentially identical to what we described above, with $u \rightarrow
Q$, apart from one complication.  (See Figure
\ref{thooftVerticesTwo}.)  The complication is, that $Q_L, Q_R$ will
also appear in a separate 't Hooft vertex, associated with $G$ (and a
phase angle $\theta^\prime$), so it is not correct that we can
redefine their relative phase, and thus alter $\theta$, without
changing anything else in the Lagrangian.   To alleviate that
difficulty, we can introduce another massless spin-$\frac{1}{2}$
fermion $S$ with different quantum numbers from $Q$ under $SU(3)$
and/or $G$.   $S$ will generally appear with different multiplicities
from $Q$ in the two vertices.   If the multiplicity pairs are linearly
independent, then we can set both $\theta$ and $\theta^\prime$ to zero
by appropriate re-definitions of $Q_L$ and $S_L$.   The simplest
possibility is for $S$ to transform under $G$ but not under $SU(3)$.

The suggested model -- an additional gauge sector $G$ with massless
(vectorlike) fermions $Q$ and $S$, which are respectively in the
fundamental and singlet representations of $SU_c(3)$ -- possesses
chiral symmetries.  If the superstrong dynamics do not break these
chiral symmetries, then all cryptohadrons will be heavy and presumably
unobservable.  However, it appears more likely that the confining
dynamics of the $G$ field spontaneously breaks these chiral
symmetries, giving rise to Nambu-Goldstone
modes, which can infest low-energy physics.  Therefore we should
investigate the pattern of symmetries and symmetry breaking.  For
first orientation, let us consider $SU_c(3)$ gauge interactions
as a perturbation.   Then $G$ possesses four cryptoquarks; $S$ and
the 3 color components of $Q$.  At the classical level, it has a
symmetry
\begin{equation}
  \label{maxsymm}
SU_L(4) \times SU_R(4) \times U_V(1) \times U_A(1) \,.
\end{equation}
The $U_A(1)$ symmetry is anomalous, and should not be included amongst
the symmetries.  (See below.)  Conventional (QCD like) confinement
triggered by the $G$ gauge fields gives rise to spontaneous breaking
according to
\begin{equation}
  \label{maxbreak}
SU_L(4) \times SU_R(4) \times U_V(1) \to
SU_V(4) \times U_V(1) \,.
\end{equation}
At this level, we obtain 15 associated Nambu-Goldstone (cryptopion) modes.

Now let us restore the $SU_c(3)$ color gauge interactions.  The
initial symmetry is reduced, at the classical level, to an $SU_V(3)$
acting on $Q$, $U_V(1) \times U_V(1)$ of $Q$ and $S$ number, and an
axial $U_A(1)$ which acts on both $Q$ and $S$, in such a way as to
avoid the $G$ anomaly.  All rotations between $Q$ and $S$ fields are
eliminated because of their different $SU_c(3)$ interactions.  And
$SU_c(3)$ also breaks the axial part of the $SU_L(3)\times SU_R(3)$
symmetry explicitly.  Further, the residual $U_A(1)$ has a color
$SU_c (3)$ anomaly.
Plausibly there will still be, as a product of confining $G$ dynamics,
an $SU_c(3)$ color octet of pseudoscalar bosons of type
$\bar {Q}_{L\alpha} Q_R^\beta$, and a similar $\mathbf{3} + \mathbf{\bar 3}$
involving $\bar S Q $ and $\bar Q S$.   They will be lighter than
typical $G$ hadrons by factors of the kind $\alpha^p_s(\Lambda_G)$,
where $\Lambda_G$ is the dynamical scale of $G$ and $p$ is a small
positive number.   Those masses could be very large by contemporary
standards, and then the phenomenological impact of such massive
particles would then be slight.

The residual $U_A(1)$ anomaly has a topological character, however,
and is associated with low-energy QCD dynamics.  In a semiclassical
framework, its physical effect arises solely from instantons.   The
would-be Nambu-Goldstone boson associated with breaking of that
approximate symmetry plays the role of a QCD axion, with its dynamical
scale set by the dynamical chiral symmetry breaking scale of $G$.

The mechanism thus exemplified is of course much more general than the
particular model we chose to analyze.
It rests on the spontaneous breaking by $G$ of a symmetry that 
is anomalous under color $SU_c(3)$, but not colored.  The 
associated would-be Nambu-Goldstone fits the conventional
phenomenological profile of a high-scale axion \cite{inv1,inv2}, but
its origin is significantly different, and perhaps more attractive.
It does not require the introduction of any fundamental
scalar field, and avoids associated mass hierarchy issues.

Our example possesses two vectorlike $U(1)$
symmetries, associated with cryptobaryon numbers.  The associated
cryptobaryons are absolutely stable, which in some cosmological
scenarios could prove problematic.  Simple variants of the model need
not have this feature.


A noteworthy connection between massless cryptoquarks and more
conventional axion schemes emerges if we expand the model just
mentioned to include two additional complex scalar fields $\phi_Q,
\phi_S$ with two Peccei-Quinn type symmetries \cite{pq} of the type
\begin{eqnarray}
Q_L ~&\rightarrow&~ e^{i\alpha} Q_L \nonumber \\
S_L ~&\rightarrow&~ e^{i\beta} S_L \nonumber \\
\phi_Q ~&\rightarrow&~ e^{i \alpha} \phi_Q \nonumber \\
\phi_S ~&\rightarrow&~ e^{i \beta} \phi_S
\end{eqnarray}
These symmetries support Yukawa couplings of the form
\begin{equation}
\Delta {\cal L} ~=~ g_Q \phi_Q \overline{Q_L} Q_R \, + \, g_S \phi_S \overline{S_L} S_R \, + \, \mathrm{h.c}.
\end{equation}
which also violate global axial $Q$ color explicitly.

If neither $\phi_Q$ nor $\phi_S$ were to develop a vacuum expectation
value, we would in essence recover the cryptoquark model just
discussed, augmented with some gratuitous scalar fields.   If one but
not the other develops a vacuum expectation value, we shall still have
the cryptoquark mechanism for $G$, in addition to a more-or-less
conventional QCD axion \cite{fw, weinb, inv1, inv2}.   If both develop
vacuum expectation values, we shall have two axion-like particles, one
more-or-less conventional, the other putting the large energy scale
$\Lambda_G$ into play (in place of $\Lambda_{\rm QCD}$), and hence
with a larger mass than might be expected, given its coupling strength
to color gluons.  (It might, for example, weigh hundreds of GeV and
be highly unstable.)  That last case is favored, since the plausible
formation of $\langle \bar {Q_L} Q_R \rangle$ and $\langle \bar {S_L}
S_R \rangle$ condensates introduces linear terms in the potentials for
$\phi_Q, \phi_S$, and induces non-zero vacuum expectation values for
them.

The basic mechanisms for addressing the strong $P, T$ problem here
described can be implemented in many ways.  In particular, they are
not inconsistent with promising ideas about unification of quantum
numbers and couplings, nor with low-energy supersymmetry.   Were a new
superstrong force to be discovered, it would be interesting to check
whether these mechanisms are operative.  Conversely, one might be
driven, from considerations within this circle of ideas, to infer the
existence of a new superstrong force.

Acknowledgement: This work is supported by the U.S. Department of
Energy under contract No. DE-FG02-05ER41360.  FW would like to thank
the Origins Project for inspiring hospitality, Lawrence Krauss, Kohei
Komada, and Richard Lebed for useful conversations, Robert Schrock and
Francesco Sannino for bringing their work to his attention.


\begin{thebibliography}{99}

\bibitem{kobayashi} M. Kobayashi, T. Maskawa {\it Prog. Theor. Phys}. {\bf 49 (2)} 652 (1973).

\bibitem{thooft} G. 't Hooft, {\it Phys. Rev. Lett}. {\bf 37} 8 (1976).

\bibitem{edmBound} C. Baker {\it et al}.  {\it Phys. Rev. Lett}. {\bf 97 (13)} 131801 (2006). 

\bibitem{quarkMasses} Reviewed by A. Manohar, C. Sachrajda: \\ pdg.lbl.gov/2014/reviews/rpp2014-rev-quark-masses.pdf

\bibitem{appelquist} T. Appelquist, M. Piai, R. Shrock {\it Phys. Lett}. {\bf B595} 442 (2004).

\bibitem{hsuSannino} S. Hsu, F. Sannino http://arxiv.org/abs/hep-ph/0408319

\bibitem{pq} R. Peccei, H. Quinn {\it Phys. Rev. Lett}. {\bf 38 (25)} 1440 (1977).

\bibitem{fw} F. Wilczek {\it Phys. Rev. Lett}. {\bf 40 (5)} 279 (1978).

\bibitem{weinb} S. Weinberg {\it Phys. Rev. Lett}. {\bf 40 (4)} 223 (1978). 

\bibitem{inv1} J. Kim {\it Phys. Rev. Lett}. {\bf 43 (2)} 103 (1979); M. Shifman, A. Vainshtein, V. Zakharov {\it Nucl. Phys}. {\bf B166} 493 (1980).

\bibitem{inv2} M. Dine, W. Fischler, M. Srednicki {\it Phys. Lett}. {\bf B104} 199 (1981); A. Zhitnitsky {\it Sov. J. Nucl. Phys}. {\bf 31} 260 (1980). 





\end{thebibliography}
\end{document}